%% file: main.tex
\documentclass[aps,pre,twocolumn,floatfix,10pt]{revtex4-2}
\usepackage{amsmath,amssymb,amsthm}
\usepackage{graphicx}
\usepackage{upquote}
\usepackage{hyperref}

\newtheorem{theorem}{Theorem}
\newtheorem{lemma}[theorem]{Lemma}
\newtheorem{proposition}[theorem]{Proposition}
\newtheorem{corollary}[theorem]{Corollary}

\theoremstyle{definition}

\theoremstyle{remark}
\newtheorem{remark}[theorem]{Remark}

\newcommand{\fr}{\operatorname{fr}}
\newcommand{\ioct}{\operatorname{ioct}}
\newcommand{\OPT}{E_0}

\begin{document}

\title{Coordination three heals, coordination four is hard:\texorpdfstring{\\}{ } the antiferromagnetic Potts model
with a costly third state}

\author{Jeehoon Choi}
\author{Suhwan Yoo}
\author{Chaeyeon Kim}
\author{Hee Young Kwon}
\author{Changyeon Won}
\email{cywon@khu.ac.kr}
\affiliation{Department of Physics, Kyung Hee University, Seoul 02447, Republic of Korea}
\date{\today}

\begin{abstract}
We study the ground states of the antiferromagnetic three-state Potts model in which the third
state carries a single-ion energy cost $D_3$, on arbitrary lattices, and show that the
difficulty of the ground-state problem is controlled by the maximum coordination $\Delta$ alone;
on planar lattices with $\Delta\le5$ it is classified completely.
On lattices with coordination at most three the third state heals frustrated bonds one for one:
every graph of maximum degree three other than $K_4$ has an independent set of sites that meets
every odd cycle and whose size equals the number $\fr$ of frustrated bonds of the Ising ground
state. This identity also follows from known results on the chromatic vertex stability of graphs;
we give a constructive proof.
Hence $\OPT(D_3)=\min(D_3,1)\,\fr$ for all $D_3$, computable in polynomial time on planar
lattices. In graph-theoretic terms, independent odd cycle transversal is solvable in polynomial
time on planar subcubic graphs.
On planar lattices with maximum coordination four, in contrast, deciding the ground-state energy is
NP-complete for every fixed $0\le D_3<2$ and polynomial for $D_3\ge2$.
The boundary is set by a local principle: a site of coordination $d$ in the third state relieves
at most $\lfloor d/2\rfloor$ frustrated bonds, so the third state is useless once
$D_3\ge\lfloor\Delta/2\rfloor$; this threshold is sharp for every $\Delta\ge3$ and reproduces
the triangular-lattice point $D_3^*=3$ of the equivalent spin-1 Blume--Capel antiferromagnet.
From a single Ising ground state the proof builds an exact ground state in
polynomial time, which for quasicrystalline, foam-like and defective lattices with thousands of
sites takes seconds to minutes.
\end{abstract}

\maketitle

%=====================================================================
\section{Introduction}
\label{sec:intro}
Geometric frustration arises whenever the interactions of a system cannot all be satisfied at
once. The smallest example is a triangle of antiferromagnetically coupled two-state spins: however
the spins are chosen, one bond remains unsatisfied [Fig.~\ref{fig:recolour}(a)]. A third local state changes this. With three
states the triangle can be satisfied completely, and on the triangular lattice the
antiferromagnetic three-state Potts model has an extensive set of unfrustrated ground states.
In many physical realisations, however, the third state is not free. It may be a non-magnetic
level split off by a crystal field, a different orbital, or a different species on the site.
The natural question is then quantitative: \emph{how much frustration does a costly third state
remove, and at which cost does it stop being worth using?}

We address this question for the antiferromagnetic three-state Potts model in which the third
state carries a single-ion energy $D_3$. The two limits are classical. For $D_3\to\infty$ only
two states remain, the model is the Ising antiferromagnet, and on planar lattices its ground state
is found exactly in polynomial time by a matching algorithm~\cite{Hadlock1975,Bieche1980,Barahona1982}.
For $D_3\to0$ the question ``is there an unfrustrated ground state'' contains planar
3-colourability, an NP-complete problem~\cite{GareyJohnsonStockmeyer1976}. The model interpolates
between an exactly solvable and a computationally hard problem, and the location of the crossover
is a property of the lattice.

The model is a three-state Potts antiferromagnet~\cite{Wu1982} and is closely related to the
spin-1 Blume--Capel antiferromagnet~\cite{Blume1966,Capel1966}, in which the
non-magnetic state $s=0$ plays the role of the third state; on the triangular lattice the
non-magnetic sublattice phase and its stability window are known~\cite{Zukovic2013}. The
two models differ only in the energy of a bond between two third-state sites
(Sec.~\ref{sec:model}); on lattices of constant coordination their energies are affinely related,
with $D_3=(z+D)/2$, whenever third-state sites are isolated, which below threshold is always the
case for coordination up to four (Theorem~\ref{thm:local}). Most studies of such models fix a lattice and map the phase diagram
numerically. Here we instead prove statements that hold on \emph{every} lattice, controlled only
by the local coordination.

Our results are as follows.
(i) \emph{Coordination three is easy.} On lattices with coordination at most three ($\Delta\le3$)
the third state heals frustrated bonds one for one, $E_0(D_3)\ge\min(D_3,1)\fr$
(Theorem~\ref{thm:sandwich} and Appendix~\ref{app:healing}), and
equality holds on every such lattice except $K_4$: every graph of maximum degree
three other than $K_4$ has an independent odd cycle transversal of size $\fr$
(Theorem~\ref{thm:healing-full}); $K_4$ can only occur as an isolated component
(Remark~\ref{rem:K4}). The ground-state energy of a planar lattice with $\Delta\le3$ is therefore
computable in polynomial time for every $D_3$. The identity $\ioct=\fr$ also follows by combining
two results of Akbari \emph{et al.} on the chromatic vertex stability number~\cite{Akbari2022}
(Sec.~\ref{sec:healing}); our proof is constructive and yields the ground state itself from a
single Ising ground state in polynomial time (Corollary~\ref{cor:one-ground-state}).
(ii) \emph{Coordination four is hard.} On planar lattices with $\Delta=4$ the ground-state
problem is NP-complete for every fixed $0\le D_3<2$ (Theorem~\ref{thm:hard4}) and polynomial for
$D_3\ge2$.
(iii) \emph{The boundary is local.} A site of coordination $d$ in the third state relieves at
most $\lfloor d/2\rfloor$ frustrated bonds, so the third state disappears from all ground states
once $D_3>\lfloor\Delta/2\rfloor$ (Theorem~\ref{thm:threshold}), and the bound is attained for every
$\Delta\ge3$ (Proposition~\ref{prop:wheel}). For regular lattices this is the Potts-normalised form
of an elementary Blume--Capel fact; its content lies in the site-resolved form for irregular and
quasiperiodic lattices and in its sharpness. It accounts for the transition points on the
triangular lattice ($D_3^*=3$) and on the face-adjacency graph of Penrose tilings ($D_3^*=2$). Below the threshold third-state sites are
isolated and, for coordination four and $1<D_3<2$, have two neighbours in each other state
(Theorem~\ref{thm:local}).
(iv) Exact ground states at coordination three are constructed and certified by matching
followed by 2-SAT, or, without possibility of failure, by the constructive procedure of
Appendix~\ref{app:healing}; we apply them to quasicrystalline, foam-like and defective lattices
with thousands of sites.
Together, (i)--(iii) classify the planar problem completely for $\Delta\le5$; only the window
$2\le D_3<\lfloor\Delta/2\rfloor$ for $\Delta\ge6$ remains open.

The paper is organised as follows. Section~\ref{sec:model} defines the model and its relation to
the Blume--Capel antiferromagnet. Section~\ref{sec:threshold} proves the threshold law and the
exact algorithm above threshold, and Sec.~\ref{sec:structure} the local structure below it.
Section~\ref{sec:healing} treats coordination three and Sec.~\ref{sec:hard4} proves hardness at
coordination four. Section~\ref{sec:lattices} applies the results to specific lattices, and
Sec.~\ref{sec:discussion} discusses the complexity landscape and open problems.
Appendix~\ref{app:numerics} describes the numerical methods and Appendix~\ref{app:healing} the
proof of the healing theorem; the Supplemental Material~\cite{SM} compares our result with the
literature on independent odd cycle transversal.

%=====================================================================
\section{Model}
\label{sec:model}
Let $G=(V,E)$ be a finite simple graph (the lattice), $d(v)$ the coordination of site $v$ and
$\Delta=\max_v d(v)$. A configuration is $s:V\to\{1,2,3\}$ with energy
\begin{equation}
E(s)=\#\{uv\in E:\ s_u=s_v\}+D_3\,\#\{v:\ s_v=3\},
\label{eq:H}
\end{equation}
with $D_3\ge 0$. In the usual notation,
\begin{equation}
E(s)=J\sum_{\langle uv\rangle}\delta_{s_us_v}+\sum_u D_{s_u}
\label{eq:Hstd}
\end{equation}
with $J=1$ and $(D_1,D_2,D_3)=(0,0,D_3)$. This is the antiferromagnetic three-state Potts model
with a single-ion (on-site) anisotropy that raises the energy of state 3. For two states the bond term is the Ising one,
$J\delta_{s_us_v}=\tfrac J2(\sigma_u\sigma_v+1)$ with $\sigma=\pm1$, so each frustrated bond costs
$J=1$ in both descriptions.
State 3 is a genuine state, not a vacancy: two adjacent sites in state 3 cost one unit.
Write $\OPT(D_3)=\min_s E(s)$ and $\fr(G)=\min_{t:V\to\{1,2\}}\#\{uv:\ t_u=t_v\}$, the number of
frustrated bonds of the Ising (two-state) ground state; $\fr(G)=|E|-\mathrm{maxcut}(G)$.
In graph-theoretic language sites are vertices, bonds are edges, the coordination is the degree,
a frustrated bond of a two-state configuration is a monochromatic edge of a 2-colouring, an Ising
ground state is a maximum cut, and $\fr(G)$ is the frustration index of $G$ with all bonds
antiferromagnetic. Graphs with $\Delta\le3$ are called subcubic, and by ``coordination three
(four)'' we mean maximum coordination $\Delta=3$ ($\Delta=4$).

\paragraph*{Relation to the Blume--Capel antiferromagnet.}
The spin-1 Blume--Capel antiferromagnet has $H_{\rm BC}=\sum_{uv\in E}\sigma_u\sigma_v-D\sum_v\sigma_v^2$
with $\sigma_v\in\{-1,0,1\}$ (exchange $|J|=1$, anisotropy $D$ in units of $|J|$).
Identify states 1, 2 with $\sigma=\pm1$ and state 3 with $\sigma=0$, and let $S=s^{-1}(3)$ and
$e(S)$ be the number of bonds inside $S$. Checking the four bond types and the two site types gives
\begin{equation}
H_{\rm BC}=2E(s)-3\,e(S)+\sum_{v\in S}\bigl(d(v)+D-2D_3\bigr)-|E|-D|V| .
\label{eq:BC}
\end{equation}
On a lattice of constant coordination $z$ and on configurations with $e(S)=0$ the two energies are
therefore affinely related when $D_3=(z+D)/2$. The genuine differences are the energy of a
$0$--$0$ bond, which is $0$ in the Blume--Capel model and $1$ here, and, on irregular lattices, the
site-dependent shift $d(v)$. On the triangular lattice ($z=6$) the Blume--Capel boundary $D=0$
between the Wannier-type and the $(1,-1,0)$ sublattice ground states~\cite{Zukovic2013}
corresponds to $D_3=3$, the threshold of Theorem~\ref{thm:threshold}. The Blume--Capel phase in
which all sites have $\sigma=0$ ($D<-3/2$) has no counterpart here because $0$--$0$ bonds are
frustrated.

\paragraph*{Reformulation.} In this notation,
\begin{equation}
\OPT(D_3)=\min_{S\subseteq V}\big[D_3|S|+e(S)+\fr(G-S)\big].
\label{eq:reform}
\end{equation}
$\OPT$ is a nondecreasing concave piecewise-linear function of $D_3$; its slope is the number of
third-state sites.

%=====================================================================
\section{Local recolouring and the universal threshold}
\label{sec:threshold}

\begin{figure}
\includegraphics[width=\columnwidth]{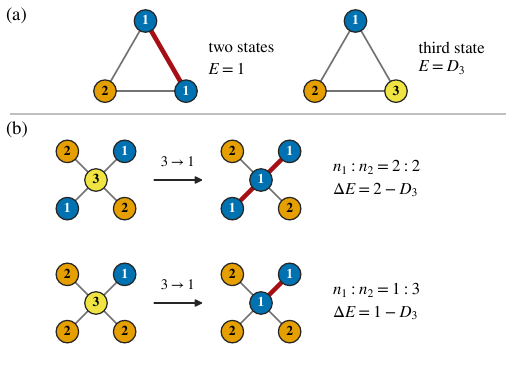}
\caption{(a) An antiferromagnetic triangle with two states always has one frustrated bond
($E=1$); with a third state it is satisfied at cost $E=D_3$. (b) Local recolouring
(Lemma~\ref{lem:recolour}) of a state-3 site of coordination $d=4$: recolouring it to the rarer
neighbouring state creates $\min(n_1,n_2)$ frustrated bonds, so $\Delta E=\min(n_1,n_2)-D_3$.
A balanced neighbourhood ($2{:}2$, top) relieves $\lfloor d/2\rfloor=2$ bonds, an unbalanced one
($1{:}3$, bottom) only one. Throughout, sites in states 1, 2, 3 are drawn blue, orange and
yellow (digits give the state; in lattices circles, squares and triangles), and frustrated
(same-state) bonds are drawn thick and red.}
\label{fig:recolour}
\end{figure}

\begin{lemma}[local recolouring]\label{lem:recolour}
Let $s_v=3$, and let $n_1,n_2,a$ be the numbers of neighbours of $v$ in states $1,2,3$.
Recolouring $v$ with the state $c\in\{1,2\}$ that is rarer among its neighbours changes the
energy by
\begin{align*}
\Delta E=\min(n_1,n_2)-a-D_3&\le\lfloor (d(v)-a)/2\rfloor-a-D_3\\
&\le\lfloor d(v)/2\rfloor-D_3 .
\end{align*}
\end{lemma}
\begin{proof}
Only bonds at $v$ change: the $a$ bonds to state-3 neighbours stop being frustrated, the
$n_c=\min(n_1,n_2)$ bonds to state-$c$ neighbours become frustrated, and the single-ion term
drops by $D_3$. The bounds follow from $\min(n_1,n_2)\le\lfloor(n_1+n_2)/2\rfloor$.
\end{proof}
A state-3 site thus relieves at most $\lfloor d(v)/2\rfloor$ frustrated bonds, and only when its
neighbours are split evenly between states 1 and 2 [Fig.~\ref{fig:recolour}(b)].

\begin{theorem}[threshold]\label{thm:threshold}
Let $W=\{v:\lfloor d(v)/2\rfloor\le D_3\}$.
(a) Every configuration can be turned, without raising its energy, into one with no state-3
site in $W$. (b) If $\lfloor d(v)/2\rfloor<D_3$, no ground state has $s_v=3$.
(c) In every ground state each state-3 site satisfies $\min(n_1,n_2)\ge a+D_3$.
In particular, if $D_3\ge\lfloor\Delta/2\rfloor$ then $\OPT(D_3)=\fr(G)$, and if
$D_3>\lfloor\Delta/2\rfloor$ no ground state uses state 3.
\end{theorem}
\begin{proof}
(a) Apply Lemma~\ref{lem:recolour} to state-3 sites in $W$ one at a time; the bound depends
only on $d(v)$, and each step removes one state-3 site. (b),(c) Lemma~\ref{lem:recolour}
applied to a ground state.
\end{proof}

\begin{figure}
\includegraphics[width=\columnwidth]{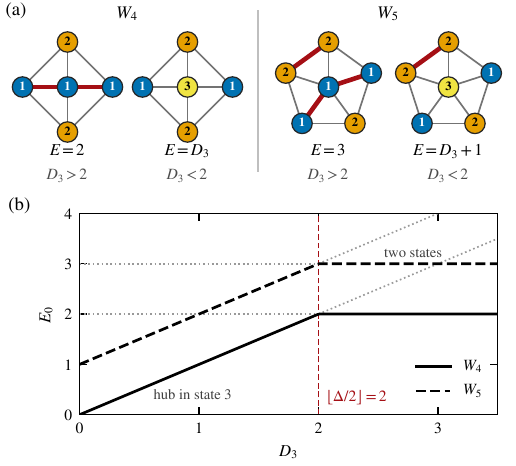}
\caption{Sharpness of the threshold (Proposition~\ref{prop:wheel}). (a) Wheels $W_4$ and $W_5$:
the best two-state configurations, with $\fr(W_\Delta)=\lceil\Delta/2\rceil=2$ and $3$
frustrated bonds, and the configurations with the hub in state 3, with energies $D_3$ and $D_3+1$
(the odd rim of $W_5$ keeps one frustrated bond); the lower labels give the range of $D_3$ in which
each configuration is the ground state. (b) Ground-state energy $E_0(D_3)$ of $W_4$
(solid) and $W_5$ (dashed); dotted lines continue the excited branches. For both wheels the
level crossing lies at $D_3=\lfloor\Delta/2\rfloor=2$.}
\label{fig:wheel}
\end{figure}

\begin{proposition}[sharpness for every $\Delta$]\label{prop:wheel}
For $\Delta\ge3$ let $W_\Delta$ be the wheel (a hub joined to a $\Delta$-cycle). For every
$D_3<\lfloor\Delta/2\rfloor$ every ground state of $W_\Delta$ uses state 3.
\end{proposition}
\begin{proof}
Each of the $\Delta$ triangles needs a frustrated bond in any two-state configuration and each
spoke lies in two triangles, so $\fr(W_\Delta)=\lceil\Delta/2\rceil$ (attained by alternating the
rim). Putting the hub in state 3 costs $D_3+\fr(C_\Delta)=D_3+(\lceil\Delta/2\rceil-\lfloor\Delta/2\rfloor)$,
which is below $\fr(W_\Delta)=\lceil\Delta/2\rceil$ exactly when $D_3<\lfloor\Delta/2\rfloor$.
\end{proof}
The two ground states of $W_4$ and $W_5$ and their level crossing at $D_3=\lfloor\Delta/2\rfloor$
are shown in Fig.~\ref{fig:wheel}.

\begin{remark}
The octahedron gives a second $\Delta=4$ example with $D_3^*=2$; the triangular lattice, where
$D_3^*=3$, is treated in Sec.~\ref{sec:lattices}.
\end{remark}

\paragraph*{Exact ground states above the threshold.} For planar $G$ and
$D_3\ge\lfloor\Delta/2\rfloor$, $\OPT=\fr(G)=|E|-\mathrm{maxcut}(G)$ follows from a planar maximum
cut, i.e.\ a minimum $T$-join between odd faces in the dual, found by minimum-weight perfect
matching~\cite{Hadlock1975,Bieche1980,Barahona1982}. The planar Ising ($D_3=\infty$) ground-state
algorithm is therefore exact for $D_3\ge3$ on subgraphs of the triangular lattice and for
$D_3\ge2$ on lattices with $\Delta\le5$.

%=====================================================================
\section{Structure of ground states below the threshold}
\label{sec:structure}
\begin{theorem}[local structure]\label{thm:local}
In every ground state and at every state-3 site:
(i) if $D_3>\lfloor(\Delta-1)/2\rfloor-1$ the state-3 sites are pairwise non-adjacent; in
particular for $\Delta\le4$ and every $D_3>0$;
(ii) if $k-1<D_3\le k$ then $\min(n_1,n_2)\ge a+k$ and $d(v)\ge 3a+2k$;
(iii) if $\Delta=4$ and $1<D_3<2$, then $d(v)=4$ and the neighbours are two in state 1 and two
in state 2;
(iv) if $\Delta=3$ and $0<D_3<1$, then $a=0$ and $n_1,n_2\ge1$; hence $d(v)\in\{2,3\}$, and
$\{n_1,n_2\}=\{1,2\}$ when $d(v)=3$.
\end{theorem}
\begin{proof}
Let $v$ be a state-3 site of a ground state and $k=\lceil D_3\rceil\ge1$. By
Theorem~\ref{thm:threshold}(c), $\min(n_1,n_2)\ge a+D_3>a+k-1$, and since the left side is an
integer, $\min(n_1,n_2)\ge a+k$. Then $d(v)\ge n_1+n_2+a\ge 3a+2k$, which is (ii).
(i) If $a\ge1$ then $\Delta\ge d(v)\ge3+2k$, i.e.\ $\lceil D_3\rceil\le\lfloor(\Delta-3)/2\rfloor
=\lfloor(\Delta-1)/2\rfloor-1$; this is excluded when $D_3>\lfloor(\Delta-1)/2\rfloor-1$. For
$\Delta\le4$ the right side is $0$, so $a=0$ for every $D_3>0$.
(iii) Here $k=2$ and $a=0$ by (i), so $n_1,n_2\ge2$; with $d(v)\le4$ this forces $n_1=n_2=2$ and
$d(v)=4$.
(iv) Here $k=1$ and $a=0$, so $n_1,n_2\ge1$ and $n_1+n_2=d(v)\le3$.
Degree-two sites can indeed carry state 3: on a triangle with one pendant edge and
$0<D_3<1$, putting a degree-two triangle vertex in state 3 is a ground state.
\end{proof}

%=====================================================================
\section{Coordination three: one-for-one healing}
\label{sec:healing}
\begin{theorem}[sandwich]\label{thm:sandwich}
If $\Delta\le3$ and $0\le D_3\le1$ then
$D_3\,\fr(G)\le\OPT(D_3)\le\min\{\fr(G),\,D_3\,\ioct(G)\}$, where $\ioct(G)$ is the minimum
size of an independent set whose removal leaves a bipartite graph.
\end{theorem}
\begin{proof}
Re-insert the state-3 sites one at a time with the rarer neighbouring state; each creates at
most one frustrated bond, so $\fr(G)\le m_{12}+|S|$ and
$E=m_{12}+e(S)+D_3|S|\ge D_3(m_{12}+|S|)$.
\end{proof}

For $\Delta\le3$ the frustrated bonds of an Ising ground state form a matching: two frustrated
bonds at one site could be removed by flipping it. We call an Ising ground state \emph{healable}
if it admits a set of pairwise non-adjacent sites containing exactly one endpoint of each
frustrated bond.

\begin{theorem}[healing criterion]\label{thm:healing}
For $\Delta\le3$ and $0<D_3<1$ the following are equivalent:
(i) $\OPT(D_3)=D_3\,\fr(G)$; (ii) $\ioct(G)=\fr(G)$; (iii) some Ising ground state is healable.
\end{theorem}
\begin{proof}
(iii)$\Rightarrow$(i): putting the chosen sites in state 3 removes every frustrated bond and
creates none, so $E=D_3\fr$, optimal by Theorem~\ref{thm:sandwich}.
(i)$\Rightarrow$(ii): in the proof of Theorem~\ref{thm:sandwich}, $E=D_3\fr$ forces
$(1-D_3)m_{12}+e(S)=0$ and $|S|=\fr$, so $S$ is independent and $G-S$ is bipartite.
(ii)$\Rightarrow$(iii): two-colour $G-S$ for an independent transversal $S$ with $|S|=\fr$ and
re-insert each site of $S$ with its rarer neighbouring state. Each creates at most one frustrated
bond, at itself, and the total is at least $\fr$, so the result is an Ising ground state whose
frustrated bonds have exactly one endpoint in the independent set $S$.
\end{proof}

\paragraph*{Certification algorithm.} (1) Ising ground state by matching; (2) choose one endpoint
of each frustrated bond by 2-SAT such that the chosen sites are pairwise non-adjacent; (3) if satisfiable, the healed configuration has energy
$D_3\fr$ and is certified optimal by Theorem~\ref{thm:sandwich}. Figure~\ref{fig:healing}(a)
illustrates the three steps; running times are given in Appendix~\ref{app:numerics}
(Table~\ref{tab:timing}, Fig.~\ref{fig:timing}).

\begin{figure}
\includegraphics[width=\columnwidth]{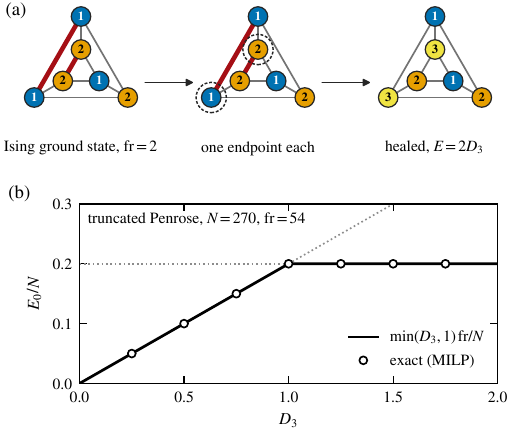}
\caption{Healing at coordination three. (a) Certification algorithm on the triangular prism:
an Ising ground state with $\fr=2$ frustrated bonds, which form a matching; one endpoint of each
frustrated bond is chosen (dashed rings) such that the chosen sites are non-adjacent, a 2-SAT
problem (here the spokes forbid choosing both top or both lower-left endpoints); putting the
chosen sites in state 3 gives a proper three-colouring with $E=2D_3=D_3\fr$, optimal for
$0\le D_3\le1$ by Theorem~\ref{thm:sandwich}. (b) Ground-state energy per site of a truncated
Penrose patch ($N=270$, $\fr=54$): the certified curve $E_0=\min(D_3,1)\,\fr$ (line) agrees with
exact mixed-integer solutions (circles) and has a single kink at $D_3=1$; dotted lines continue
the two branches.}
\label{fig:healing}
\end{figure}

The certification algorithm succeeds whenever the chosen Ising ground state is healable. Not every
Ising ground state is healable (an 18-site example is given in Appendix~\ref{app:numerics}), but
every graph other than $K_4$ has a healable one:

\begin{theorem}[healing]\label{thm:healing-full}
Let $G$ be a graph of maximum degree at most three with no connected component isomorphic to
$K_4$. Then $\ioct(G)=\fr(G)$. Consequently $\OPT(D_3)=\min(D_3,1)\,\fr(G)$ for every $D_3\ge0$.
\end{theorem}

\begin{remark}\label{rem:K4}
In a graph of maximum degree three, a subgraph isomorphic to $K_4$ uses all three bonds of each
of its four sites, so it is always an entire connected component. The hypothesis of
Theorem~\ref{thm:healing-full} therefore holds automatically for every connected lattice with
more than four sites; for such lattices the theorem has no exception. The exception is real only
for an isolated $K_4$, where $\OPT(D_3)=\min(2,1+D_3)>\min(D_3,1)\,\fr$ for $0<D_3<1$.
\end{remark}

The proof (Appendix~\ref{app:healing}) takes a set $S$ of non-adjacent sites that is \emph{tight},
$|S|+\fr(G-S)=\fr(G)$, and of maximum size. If $G-S$ were still frustrated, its frustrated
components would be odd cycles each of whose sites has exactly one neighbour in $S$; exchanging a
cycle site with that neighbour produces a new odd cycle in another component of $G-S$, and the
resulting chain of exchanges would never end in a finite graph. Because maximality is used only
for the sets met along such a chain, the argument is constructive: greedy addition of sites,
supplemented by these exchanges when it gets stuck, builds the transversal
(Proposition~\ref{prop:constructive}), and one Ising ground state suffices as input
(Corollary~\ref{cor:one-ground-state}). Pure greedy addition is not enough; it gets stuck on some
graphs with 12 sites.

\paragraph*{Relation to earlier results.}
For $\Delta\le3$, Choi, Nakajima and Rim showed that the minimum number of vertices whose removal
leaves a bipartite graph equals $\fr(G)$~\cite{Choi1989}; Theorem~\ref{thm:healing-full}
requires in addition the removed vertices to be independent. In the language of chromatic
stability, for a graph with chromatic number $\chi=3$ the quantities $\fr(G)$, the minimum odd cycle
transversal and $\ioct(G)$ are the minimum numbers of edges, of vertices and of independent
vertices whose removal lowers $\chi$ by one. Akbari \emph{et al.} proved that the two vertex
numbers coincide whenever $\chi\in\{\Delta,\Delta+1\}$, and that the vertex and edge numbers
coincide whenever $\chi>\Delta/2+1$~\cite{Akbari2022}. For $\Delta=3$ the second result is that of
Choi, Nakajima and Rim, and together the two imply Theorem~\ref{thm:healing-full}. Their proof of
the first result is not an algorithm: it starts from a minimum deletion set chosen, among all of
them, with the fewest internal edges, together with an extremal colouring of the remainder, and
argues by induction on the size of the deletion set. The chain-and-exchange proof of
Appendix~\ref{app:healing} uses extremality only for explicitly given sets and is therefore
algorithmic (Proposition~\ref{prop:constructive}, Corollary~\ref{cor:one-ground-state}). The
condition $\chi>\Delta/2+1$ is the condition $\lfloor\Delta/(\chi-1)\rfloor\le1$ of the
re-insertion bound used in Theorem~\ref{thm:sandwich}. $K_4$ is also the
exception in Brooks' theorem~\cite{Brooks1941}. Chen, Li and Wang bounded the frustration index of
signed subcubic graphs, giving $\fr(G)\le|V|/3\le\alpha(G)$ for 2-edge-connected $G\ne K_4$ up to
finitely many small exceptions~\cite{ChenLiWang2025}, consistent with the theorem. The theorem
was checked exhaustively for all 27\,523 connected graphs with $\Delta\le3$ on at most 12
vertices and by exact optimisation on 1\,637 larger graphs with up to 100 vertices
(Appendix~\ref{app:numerics}).

%=====================================================================
\section{Coordination four: hardness below the threshold}
\label{sec:hard4}
\begin{theorem}[hardness at $\Delta=4$]\label{thm:hard4}
For every fixed rational $0<D_3<2$ it is NP-complete to decide, given a planar graph $G$ of
maximum degree four and a number $K$, whether $\OPT(D_3)\le K$.
\end{theorem}
Together with Theorem~\ref{thm:threshold} and planar 3-colourability at $D_3=0$
\cite{GareyJohnsonStockmeyer1976}, the planar $\Delta=4$ problem is NP-complete for $0\le D_3<2$
and polynomial for $D_3\ge2$.

\begin{proof}
We use the graph $G_\varphi$ with which Choi, Nakajima and Rim showed that vertex bipartization
is NP-complete on planar graphs of maximum degree four~\cite{Choi1989}, starting from
planar 3-SAT with exactly three literals per clause~\cite{Lichtenstein1982}. For a variable with
$n_i$ occurrences, $G_\varphi$ contains a cycle $b_1b_2\cdots b_{4n_i}$ with a triangle erected on
each cycle bond; the $b$'s alternately represent $x_i$ and $\bar x_i$. Each clause is a path
$w_1w_2w_3w_4$ of four new sites which, through the tips of two adjacent triangles at each of its
three literals, closes a single cycle of length 13 that passes through one $b$ of each literal's
type.
$G_\varphi$ has maximum degree four, $28m$ sites for $m$ clauses, and $12m$ variable
triangles. It is planar provided the occurrence blocks around each variable cycle follow the
rotation of that variable in a planar embedding of the variable--clause incidence graph and the
literals of each clause are ordered along $w_1\cdots w_4$ opposite to the rotation at the clause
(we verified planarity for 307 planar formulas with this rule; with the same orientation at
variables and clauses only 67 of them were planar). Choi \emph{et al.} show that $\varphi$ is satisfiable iff removing the $6m$ $b$'s of the
true literal types leaves a bipartite graph; these $b$'s are pairwise non-adjacent.

Set $K=6mD_3$. If $\varphi$ is satisfiable, put these $6m$ sites in state 3 and two-colour the rest;
the energy is $6mD_3$. Conversely, distribute the energy of any configuration over the variable
triangles: each triangle receives its own frustrated bonds, the full $D_3$ of a state-3 tip and
$D_3/2$ of each state-3 base site (a base site lies in two triangles). The triangles are
bond-disjoint, so the charges sum to at most $E(s)$. A triangle without state 3 has a frustrated
bond, charge $\ge1$; one with a state-3 tip has charge $\ge D_3$; otherwise charge $\ge D_3/2$.
Since $D_3<2$ every charge is $\ge D_3/2$, so $E(s)\ge6mD_3$, with equality only if every triangle
contains exactly one state-3 site, a base site, and nothing else costs energy. Then the state-3
sites alternate along each variable cycle, fixing a truth assignment, the configuration is a
proper colouring, and every odd clause cycle contains a state-3 site, i.e.\ a true literal.
\end{proof}

Exact solutions of the reduction confirm the gap: for the unsatisfiable formula consisting of all
eight clauses on three variables ($224$ sites; this instance is not planar, but the bound
$E\ge6mD_3$ does not use planarity) we find $\OPT-6mD_3=\min(D_3,2-D_3)$ at $D_3=0.5,1.5,1.9$,
while satisfiable formulas give equality (Appendix~\ref{app:numerics}). The gap closes as
$D_3\to2$, consistent with the threshold. The same instances show hardness for $0<D_3<2$ at every
$\Delta\ge4$.

%=====================================================================
\section{Lattices}
\label{sec:lattices}
\begin{figure}
\includegraphics[width=\columnwidth]{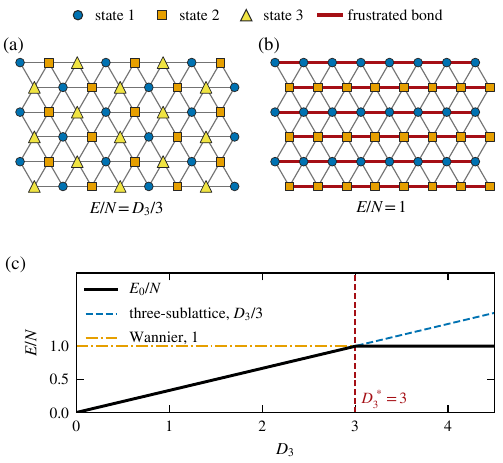}
\caption{Triangular lattice ($\Delta=6$). (a) Three-sublattice state: a proper three-colouring
with one third of the sites in state 3, $E=ND_3/3$. (b) Two-state stripe configuration, one of
the Wannier ground states, with exactly one frustrated bond per triangle, $E=\fr=N$. (c) Energy
per site of the two states and the ground-state energy $E_0/N=\min(D_3/3,1)$; the level crossing
is at $D_3^*=3=\lfloor6/2\rfloor$.}
\label{fig:triangular}
\end{figure}

\paragraph*{Triangular lattice ($\Delta=6$).} Every two-state configuration has at least one
frustrated bond per triangle. Each bond lies in two triangles, so $\fr\ge N$ for $N$ sites (with
$2N$ triangles), and the Wannier states attain $N$~\cite{Wannier1950}. The three-sublattice
configuration has energy $ND_3/3$. The ground state therefore changes at $D_3^*=3=\lfloor6/2\rfloor$
by a level crossing, in agreement with the Blume--Capel
result~\cite{Zukovic2013} ($D=0$ in Eq.~\eqref{eq:BC}). The triangular lattice is thus an extremal lattice for
Theorem~\ref{thm:threshold}. The two competing states are shown in Fig.~\ref{fig:triangular}.

\begin{figure*}
\includegraphics[width=\textwidth]{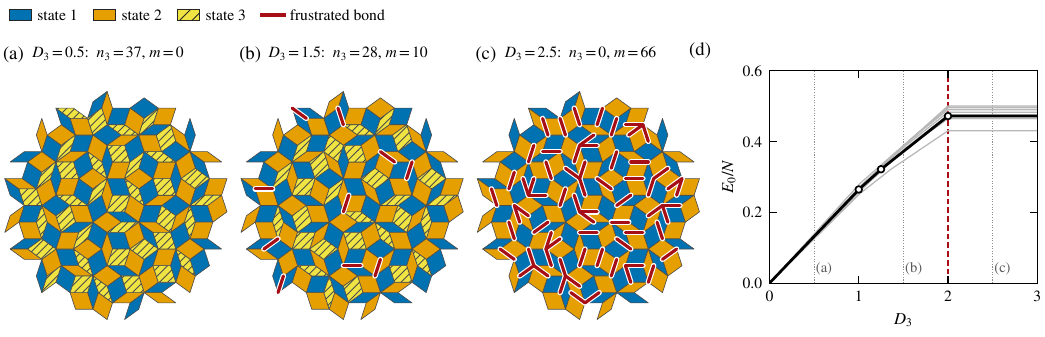}
\caption{Penrose face-adjacency graph ($\Delta=4$), patch of 140 rhombi. (a)--(c) Exact ground
states at $D_3=0.5$, $1.5$ and $2.5$; rhombi are coloured by their state (state 3 hatched) and
edges shared by two rhombi in the same state (frustrated bonds) are drawn in red; $n_3$ is the
number of state-3 rhombi and $m$ the number of frustrated bonds. At $D_3=1.5$ every state-3
rhombus has two neighbours in each of the other states; at $D_3=2.5$ the state is an Ising ground
state with $m=\fr=66$. (d) Exact ground-state energy per rhombus $E_0/N$ of this patch (black,
kinks at $D_3=1$, $5/4$ and $2$ marked by circles) and of nine further patches with 97--248
rhombi (grey). The dashed line marks the threshold $D_3^*=\lfloor4/2\rfloor=2$; dotted lines mark
the values of panels (a)--(c).}
\label{fig:penrose}
\end{figure*}

\paragraph*{Penrose face-adjacency graph ($\Delta=4$).} Taking the rhombi of a Penrose P3 tiling
as sites and edge-sharing rhombi as neighbours gives a four-coordinated planar graph. Odd
cycles surround the tiling vertices of odd degree (3, 5, 7), so the graph is frustrated. We
computed exact ground-state energy curves of patches by mixed-integer programming, together with
the exact lower envelope over $D_3$ (Appendix~\ref{app:numerics}). All state-3 rhombi disappear
together at $D_3^*=2$. For a 140-rhombus patch at $D_3=1.5$ the ground state has $n_3=28$ state-3
rhombi and $m=10$ frustrated bonds, and $\fr=66=10+2\times28$: each state-3 rhombus relieves
exactly $\lfloor4/2\rfloor=2$ frustrated bonds. At $D_3=1.5$ all 28 have
two neighbours in each of the other states, as required by Theorem~\ref{thm:local}(iii).
The two principal kinks are those allowed by Theorem~\ref{thm:local}: below $D_3=1$ state-3
rhombi may have a $1{:}3$ neighbourhood, above it only $2{:}2$, and at $D_3=2$ all state-3 rhombi
disappear [Fig.~\ref{fig:penrose}]. Between them the exact curves of ten patches (97--248 rhombi;
four patch radii, with three phason offsets for the three smaller radii) show one to three
further kinks at patch-dependent positions ($4/3$, $5/4$, $7/5$, $3/2$, $5/3$). They are not repairs
confined to the patch edge: most of the rhombi that change lie two or more steps from the edge,
and the changes reach the patch centre. Their energy gain relative to the envelope formed by the
two principal segments, however, fluctuates strongly ($0.004$--$0.024$ per rhombus, the smallest
value for the largest patch). Whether these kinks survive in the thermodynamic limit cannot be
decided from open patches and requires periodic approximants.

\begin{figure*}
\includegraphics[width=\textwidth]{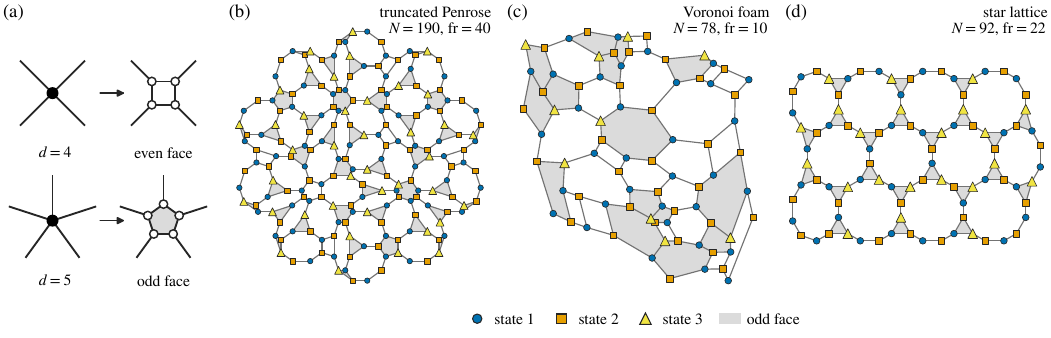}
\caption{Truncated lattices ($\Delta=3$). (a) Truncation replaces a vertex of degree $d$ by a
$d$-gon; the new face is odd (grey) exactly when $d$ is odd, while faces of the parent tiling
double in length. (b)--(d) Certified ground states for $0<D_3<1$ obtained by the
matching-plus-2-SAT algorithm: (b) truncated Penrose tiling ($N=190$ sites, $\fr=40$),
(c) Voronoi foam of 45 random points ($N=78$, $\fr=10$), (d) star (3.12.12) lattice ($N=92$,
$\fr=22$). Odd faces are shaded. In each case $\fr$ pairwise non-adjacent sites are in state 3
(triangles), no frustrated bond remains, and $E_0=D_3\fr$.}
\label{fig:truncation}
\end{figure*}

\paragraph*{Truncated lattices ($\Delta=3$).} Replacing every vertex of degree $d$ of a planar
tiling by a $d$-gon (truncation) produces a three-coordinated lattice. Faces of the parent tiling
double in length and stay even, so \emph{the odd faces of the truncated lattice sit exactly at
the odd-degree vertices of the parent}. Truncating the honeycomb gives the star (3.12.12)
lattice, truncating the rhombille lattice gives triangles, hexagons and octagons, and truncating a
Penrose tiling gives a quasiperiodic lattice whose odd faces are triangles, pentagons and
heptagons. The truncated
Penrose lattice carries its frustration at the same places as the Penrose face-adjacency graph,
so the two allow a direct comparison between coordination four and three.
Figure~\ref{fig:truncation} shows the construction and certified ground states of three such
lattices.

\paragraph*{Certified ground states at coordination three.} For truncated Penrose patches,
the star lattice, the truncated rhombille lattice, honeycombs with 5--7 defects and random
Voronoi (foam) networks, the matching-plus-2-SAT algorithm of Sec.~\ref{sec:healing} succeeded on
the first Ising ground state in every case, and the constructive procedure of
Corollary~\ref{cor:one-ground-state} gave the same energies. The resulting energy curve is
$E_0(D_3)=\min(D_3,1)\fr$, with a single kink at $D_3=1$ [Fig.~\ref{fig:healing}(b)]. This contrasts with the several kinks of
the four-coordinated Penrose graph. Patches with $3$--$4\times10^3$ sites were solved and
certified in 3--35\,s (Table~\ref{tab:timing}, Fig.~\ref{fig:timing}).

%=====================================================================
\section{Discussion}
\label{sec:discussion}
\paragraph*{Complexity landscape.} On planar lattices the ground-state problem is polynomial for
$D_3\ge\lfloor\Delta/2\rfloor$ (Theorem~\ref{thm:threshold}). For $\Delta\le3$ it is polynomial for
every $D_3$ (Theorem~\ref{thm:healing-full}; at $D_3=0$ this is Brooks' theorem~\cite{Brooks1941}).
For $\Delta=4,5$ it is NP-complete for $0\le D_3<2$ (planar 3-colourability~\cite{GareyJohnsonStockmeyer1976}
and Theorem~\ref{thm:hard4}) and polynomial for $D_3\ge2$. Coordination three and four thus sit on
opposite sides, and the threshold $\lfloor\Delta/2\rfloor$ is the exact complexity boundary for
$\Delta\le5$. At coordination three healing makes the problem easy; at coordination four it does
not, because a state-3 site can relieve two bonds and the choice of which sites to use carries
information. The independence of state-3 sites, forced below the threshold by
Theorem~\ref{thm:local}, also has a cost at coordination four: there are infinitely many graphs
with $\Delta=4$ and $\chi=3$ for which the smallest independent set whose removal leaves a
bipartite graph is larger than the smallest odd cycle transversal~\cite{Knor2022}. More
generally, for $3\le\Delta\le10$ the condition $\chi\ge\Delta$ under which an optimal deletion set
can be taken independent cannot be weakened~\cite{Cambie2024}.

\paragraph*{More states.} The same structure holds for $q$ states of which one is costly: if
$\Delta\le q$ and no component is $K_{q+1}$, then $\OPT(D_q)=\min(D_q,1)\,\fr_{q-1}(G)$, where
$\fr_{q-1}$ is the minimum number of monochromatic bonds of a $(q-1)$-colouring. The lower bound is
the re-insertion argument of Theorem~\ref{thm:sandwich}, and the matching independent transversal
exists by Ref.~\cite{Akbari2022}. For $q=4$, however, computing $\fr_3$ is NP-hard already on planar
lattices, since deciding $\fr_3=0$ is planar 3-colourability at maximum degree
four~\cite{GareyJohnsonStockmeyer1976}; the structural statement generalises but the polynomial
algorithm does not.

\paragraph*{A cost is not a field.} Barahona showed that a magnetic field turns the planar Ising
ground-state problem from polynomial into NP-hard~\cite{Barahona1982}. The third-state cost looks
like a field on one state, but unlike a field it preserves the symmetry between states 1 and 2.
Standard hardness constructions use a global reference for ``true'' and ``false'', which a
symmetric cost does not provide. The reduction of Theorem~\ref{thm:hard4} avoids this: truth is
the choice of one of the two alternating state-3 patterns on each variable cycle, a local
symmetry-breaking that the cost itself enforces. When the problem is formulated as a planar
Boolean constraint satisfaction problem (CSP), its symmetric structure places it close to the
tractable side of the planar CSP classification~\cite{KazdaKolmogorovRolinek2018,FullaZivny2017}.

\paragraph*{Independent odd cycle transversal.} Theorem~\ref{thm:healing-full}, or equivalently
the results of Ref.~\cite{Akbari2022} published in 2022, together with planar maximum
cut~\cite{Hadlock1975}, implies that independent odd cycle transversal is solvable in polynomial
time on planar subcubic graphs; Corollary~\ref{cor:one-ground-state} also constructs the
transversal. This differs from a statement made in the text following Theorem~11 of Ref.~\cite{Johnson2025}, that
the problem is NP-complete on planar subcubic graphs. That theorem itself concerns hardness on
subcubic graphs without planarity and is consistent with our result; the difference concerns only
the planarity of the constructed graphs. See the Supplemental Material~\cite{SM} for the detailed
comparison, including explicit instances.

\paragraph*{Open problems.} (1) The constructive procedure (Proposition~\ref{prop:constructive},
Corollary~\ref{cor:one-ground-state}) needed at most three chain stages in all our tests; is the
chain length bounded by a constant? (2) Decide the complexity of the window
$2\le D_3<\lfloor\Delta/2\rfloor$ for $\Delta\ge6$, e.g.\ on subgraphs of the triangular lattice.
(3) Do the intermediate kinks of the Penrose face-adjacency graph survive in the thermodynamic
limit? (4) Finite temperature: the rigid local structure of
Theorem~\ref{thm:local} suggests constrained low-energy excitations below the threshold.

\section*{Data availability}
The code and data that support the findings of this article are openly available in the
repository AF\_Potts\_Model~\cite{AFPottsRepo}. It contains the exact mixed-integer solver, the
planar matching and certification algorithms, the constructive healing procedure, the scripts that
produce every numerical result, table and figure of this article and of the Supplemental Material,
and the raw data: the exhaustive enumeration and greedy-augmentation logs, the timing data of
Table~\ref{tab:timing}, the exact energy curves of the Penrose patches, and the instances of the
Supplemental Material.

\begin{acknowledgments}
This research was supported by the National Research Foundation (NRF) of Korea funded by the
Korean Government (NRF-2023R1A2C1006050). The authors used Claude (Anthropic) to assist with
writing code for the numerical computations and figures and with editing the manuscript; all
results were checked by the authors, who take full responsibility for the content.
\end{acknowledgments}

\appendix
\section{Numerical methods}
\label{app:numerics}
\begin{figure}
\includegraphics[width=\columnwidth]{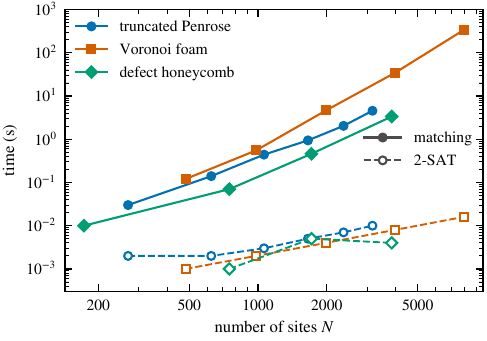}
\caption{Wall-clock time of the certification algorithm versus the number of sites $N$ for
truncated Penrose patches, Voronoi foams and defect honeycombs. Filled symbols and solid lines:
matching step (Ising ground state as a minimum $T$-join); open symbols and dashed lines: 2-SAT
step. 2-SAT times below the 1\,ms resolution are omitted. Data as in Table~\ref{tab:timing}.}
\label{fig:timing}
\end{figure}

\paragraph*{Exact solver.} For moderate sizes we minimise Eq.~\eqref{eq:H} exactly as a mixed-integer
linear programme: binary variables $x_{v,c}$ ($c=1,2,3$, one colour per site) and, for each bond
and colour, a variable $y_{uv,c}\ge x_{u,c}+x_{v,c}-1$ counting monochromatic bonds (the $c=3$
terms can be dropped to omit the cost of 3--3 bonds). The programme is solved with
HiGHS~\cite{HuangfuHall2018} through SciPy.
\paragraph*{Energy curves.} Since $\OPT(D_3)$ is concave and piecewise linear, its exact lower
envelope on an interval is obtained by bisection on the intersections of the supporting lines
$m+D_3 n_3$ returned by the solver, $m$ being the number of frustrated bonds and $n_3$ the number of
state-3 sites of the returned configuration; a segment is accepted when the solver at the intersection
point returns no line below the two neighbouring ones.
\paragraph*{Ising ground states.} For planar lattices $\fr$ is computed as a minimum $T$-join between
the odd faces of a planar embedding, i.e.\ a minimum-weight perfect matching on the odd faces with
dual shortest-path distances~\cite{Hadlock1975,Barahona1982}; we use the blossom implementation
of NetworkX~\cite{Hagberg2008}. The 2-SAT step of the certification algorithm has one Boolean variable per frustrated
bond (which endpoint takes state 3) and one clause per bond joining two such endpoints.
\paragraph*{Timing.} Table~\ref{tab:timing} lists the certification algorithm on three families of
coordination-three lattices. Every instance was healed on the first Ising ground state. The
matching step dominates; the 2-SAT step takes milliseconds. Figure~\ref{fig:timing} shows
how both steps scale with the number of sites. The constructive procedure of
Corollary~\ref{cor:one-ground-state}, started from the same matching ground state, took
$0.08$--$1.1$\,s for lattices with $1.7$--$4\times10^3$ sites (truncated Penrose, foam and defect
honeycomb); on two truncated Penrose patches it met a stuck set, which a single exchange resolved.

% BEGIN generated timing table (make_figs.py)
\begin{table}
\caption{Certification algorithm on coordination-three lattices: number of sites $N$,
Ising frustration $\fr$, wall-clock times of the matching step (Ising ground state as a
minimum $T$-join) and of the 2-SAT step (choice of endpoints), and whether the first Ising
ground state was healed, which certifies $E_0(D_3)=\min(D_3,1)\fr$. Python/NetworkX implementation.}
\label{tab:timing}
\begin{ruledtabular}
\begin{tabular}{lrrrrc}
lattice & $N$ & $\fr$ & matching & 2-SAT & healed\\
 & & & (s) & (ms) & \\
\hline
truncated Penrose & 270 & 54 & 0.03 & 2 & yes\\
truncated Penrose & 624 & 122 & 0.14 & 2 & yes\\
truncated Penrose & 1062 & 226 & 0.44 & 3 & yes\\
truncated Penrose & 1654 & 334 & 0.94 & 5 & yes\\
truncated Penrose & 2368 & 490 & 2.04 & 7 & yes\\
truncated Penrose & 3168 & 674 & 4.55 & 10 & yes\\
Voronoi foam & 482 & 71 & 0.12 & 1 & yes\\
Voronoi foam & 980 & 124 & 0.55 & 2 & yes\\
Voronoi foam & 1982 & 269 & 4.62 & 4 & yes\\
Voronoi foam & 3982 & 518 & 34.52 & 8 & yes\\
Voronoi foam & 7974 & 1059 & 340.28 & 16 & yes\\
defect honeycomb & 173 & 15 & 0.01 & $<1$ & yes\\
defect honeycomb & 749 & 50 & 0.07 & 1 & yes\\
defect honeycomb & 1713 & 95 & 0.46 & 5 & yes\\
defect honeycomb & 3847 & 208 & 3.35 & 4 & yes\\
\end{tabular}
\end{ruledtabular}
\end{table}
% END generated timing table

\paragraph*{Exhaustive enumeration.} Every connected graph has a vertex whose removal leaves it
connected, so all connected graphs with $\Delta\le3$ on $n+1$ vertices arise from those on $n$
vertices by adding a vertex joined to one, two or three vertices of degree at most two.
Isomorphism classes were separated by Weisfeiler--Lehman hashes followed by exact isomorphism
tests. This gives 2, 6, 10, 29, 64, 194, 531, 1733, 5524, 19430 graphs for $n=3,\dots,12$ (27\,523
in total); for each, some maximum cut was checked to admit an independent transversal of its
frustrated matching. Only $K_4$ fails.
\paragraph*{Strong version.} The statement that \emph{every} Ising ground state is healable is
false: an 18-site, 27-bond graph with $\fr=5$ has 13 Ising ground states, one of which is not
healable. In it a frustrated bond $uv$ joins two triangles whose other frustrated bonds have both
endpoints adjacent to $u$ and to $v$, respectively; whichever endpoint of $uv$ is chosen, it is
adjacent to both endpoints of another frustrated bond.
\paragraph*{Further checks.} Random connected subcubic graphs with 12--20 vertices: greedy
augmentation (Lemma~\ref{lem:aug}) reached $\fr(G-S)=0$ in all 972 runs; exhaustively, stuck
tight sets with $\fr(G-S)\ge1$ first occur at 12 vertices, where the exchange step of
Proposition~\ref{prop:constructive} resolves them. Exact optimisation on 251 further graphs with
14--60 vertices (cubic, triangle-rich, truncated, and the subcubic reduction graphs of
Ref.~\cite{Johnson2025}) and on 1\,386 graphs with 16--100 vertices (729 of them planar) gave
$\ioct=\fr$ in every case. For Theorem~\ref{thm:hard4}, 307 planar 3-SAT formulas
gave planar $G_\varphi$ under the ordering rule, and the energy criterion was confirmed by exact
optimisation on satisfiable and unsatisfiable instances.
\paragraph*{Penrose patches.} Patches of the Penrose face-adjacency graph were generated by the
de Bruijn pentagrid method~\cite{deBruijn1981} with random phason offsets; exact energy curves
were computed on $0\le D_3\le3$ for patches of 97--248 rhombi.

\input{healing_proof}

\bibliography{refs}

\makeatletter
% Supplemental Material, compiled separately from supplementary.tex, is appended here.
% Each page of supplementary.pdf is placed full-size on its own page with graphicx only
% (the package pdfpages conflicts with the REVTeX output routine).
\clearpage
\onecolumngrid
\newcount\supppage
\newcount\supppages
\pdfximage{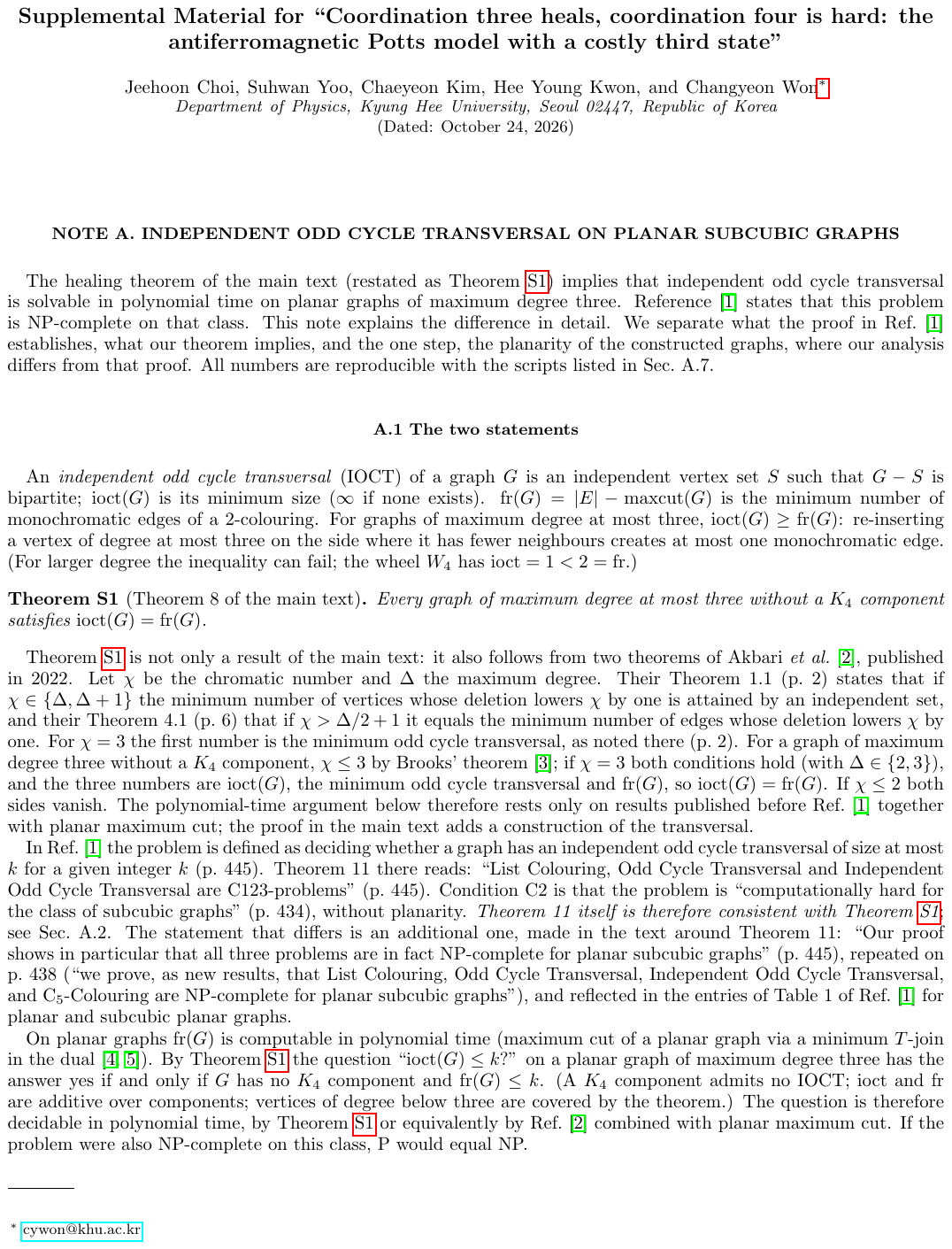}\supppages=\pdflastximagepages
\supppage=0
\loop\ifnum\supppage<\supppages
  \advance\supppage by 1
  \clearpage\thispagestyle{empty}%
  \noindent\begingroup\setlength{\unitlength}{1pt}%
  \begin{picture}(0,0)%
    \put(\strip@pt\dimexpr-1in-\hoffset-\oddsidemargin\relax,%
         \strip@pt\dimexpr1in+\voffset+\topmargin+\headheight+\headsep-\paperheight\relax)%
        {\includegraphics[page=\the\supppage,width=\paperwidth,height=\paperheight]{supplementary.pdf}}%
  \end{picture}\endgroup
\repeat
\makeatother

\end{document}

%% file: healing_proof.tex
% Proof of the healing theorem (Theorem thm:healing-full, stated in Sec. healing of main.tex).
% \input after \appendix in main.tex.  Graphs simple, maximum degree <= 3.

\section{Proof of the healing theorem}
\label{app:healing}
We give a constructive proof of Theorem~\ref{thm:healing-full}. The identity $\ioct(G)=\fr(G)$
also follows from two results of Ref.~\cite{Akbari2022} (Sec.~\ref{sec:healing}); the proof
below differs in that it uses extremality only for explicitly given sets, which makes it an
algorithm (Proposition~\ref{prop:constructive}).
We use graph-theoretic language: vertices, edges, degree, and 2-colourings, a frustrated bond being
a monochromatic edge. For $X\subseteq V$ let $\varphi(X)=|X|+\fr(G-X)$.

\emph{Proof of Theorem~\ref{thm:healing-full}.}
\paragraph*{Preliminaries.}
(F1) For every $X\subseteq V$, $\varphi(X):=|X|+\fr(G-X)\ge\fr(G)$: re-insert the vertices of $X$
one at a time, each on the side opposite to the majority of its already present neighbours;
since degrees are at most three, each insertion creates at most one monochromatic edge.
In particular every odd cycle transversal has at least $\fr(G)$ vertices, so it suffices to
find an independent $S$ with $\varphi(S)=\fr(G)$ and $\fr(G-S)=0$.
An independent set $S$ is \emph{tight} if $\varphi(S)=\fr(G)$; the empty set is tight.
Both $\fr$ and $\ioct$ are additive over connected components, so we may assume $G$
connected and $G\ne K_4$. Fix a tight set $S$ of maximum size and suppose, for a
contradiction, that $k:=\fr(G-S)\ge1$.

\begin{lemma}[augmentation]\label{lem:aug}
Let $S$ be tight and let $v$ be an endpoint of a monochromatic edge in some optimal
2-colouring of $G-S$. If $v$ has no neighbour in $S$, then $S\cup\{v\}$ is tight.
\end{lemma}
\begin{proof}
Deleting $v$ from that colouring removes at least one monochromatic edge, so
$\fr(G-S-v)\le k-1$ and $\varphi(S\cup\{v\})\le\varphi(S)$; equality holds by (F1), and
$S\cup\{v\}$ is independent.
\end{proof}

\begin{lemma}[stuck structure]\label{lem:stuck}
Let $T$ be a tight set of maximum size. Every component $K$ of $G-T$ with $\fr(K)\ge1$ is an odd
cycle, and each vertex of $K$ has degree three in $G$ and exactly one neighbour in $T$.
\end{lemma}
\begin{proof}
Take an optimal colouring of $G-T$ and a monochromatic edge $uv$ in $K$. By
Lemma~\ref{lem:aug} and maximality $v$ has a neighbour in $T$, so $\deg_K v\le2$. If
$\deg_K v=1$, recolouring $v$ improves the colouring. Hence $v$ has a second neighbour $v_1$
in $K$, of the other colour (otherwise recolouring $v$ gains two). Recolouring $v$ gives another
optimal colouring in which $v v_1$ is the monochromatic edge, so $v_1$ has the same properties.
Iterating gives a walk $v=v_0,v_1,v_2,\dots$ in which each $v_i$ has degree exactly two in $K$
and $v_{i+1}$ is the neighbour of $v_i$ other than $v_{i-1}$ ($v_{-1}=u$). By symmetry $u$ also has
degree two. Let $v_m$ be the first vertex of the walk whose successor was already visited or is
$u$; since all visited vertices have both neighbours accounted for, the successor must be $u$, so
$u,v_0,\dots,v_m$ is a cycle in which every vertex has degree two in $K$. As $K$ is connected,
$K$ is this cycle, odd because $\fr(K)\ge1$. Each vertex of $K$ has a neighbour in $T$ and degree at most
three, hence exactly one such neighbour and degree three.
\end{proof}

\begin{lemma}[exchange]\label{lem:exchange}
Let $T$ be a tight set of maximum size, $Y$ an odd-cycle component of $G-T$, $y\in Y$ and $t$ the
neighbour of $y$ in $T$. Then
(a) $\deg t=3$ and $y$ is the only neighbour of $t$ on $Y$;
(b) the other two neighbours of $t$ are the endpoints of a component $Q$ of $G-T$ which is a path
with an even number of vertices;
(c) $T'=T-t+y$ is a tight set of maximum size and $Q+t$ is an odd-cycle component of $G-T'$.
In particular distinct vertices of $Y$ have distinct neighbours in $T$.
\end{lemma}
\begin{proof}
$T'$ is independent, since the neighbours of $y$ other than $t$ lie on $Y$. Removing $y$ turns $Y$
into a path, so $\fr(G-T-y)=k-1$; adding back $t$, which has at most two neighbours in
$G-T'$, creates at most one monochromatic edge. Thus $\fr(G-T')\le k$, and by (F1) $T'$ is tight;
it has maximum size, so Lemma~\ref{lem:stuck} applies to $T'$.
If $t$ had a neighbour on another odd-cycle component $Y_2$ of $G-T$, the component of $G-T'$
containing $Y_2$ would be frustrated but contain a vertex of degree three, contradicting
Lemma~\ref{lem:stuck}. Hence the component $D$ of $t$ in $G-T'$ satisfies
$\fr(G-T')=(k-1)+\fr(D)$, so $\fr(D)=1$ and, by Lemma~\ref{lem:stuck}, $D$ is an odd cycle all of
whose vertices have degree two in $G-T'$. Thus $t$ has exactly two neighbours $t_1,t_2$ in
$G-T'$, $\deg t=3$, and $W=D-t$ is a component of $G-T-y$ which is a $t_1$--$t_2$ path with an
even number of vertices.
If $W=Y-y$, then $t$ is adjacent to both neighbours $y^\pm$ of $y$ on $Y$, and $t$ is the unique
$T$-neighbour of $y^+$. Applying the same argument to $y^+$ shows that the two neighbours of $t$
other than $y^+$ are the neighbours of $y^+$ on $Y$, i.e.\ $\{y,y^-\}=\{y,y^{++}\}$. Hence $|Y|=3$
and $Y\cup\{t\}$ induces a $K_4$ whose vertices all have degree three, so $G=K_4$, excluded.
Otherwise $W$ is a component of $G-T$ other than $Y$; it is not an odd cycle, so $W=Q$ is a
path component as claimed, and (a)--(c) follow.
\end{proof}

\paragraph*{Chains.}
Let $Z$ be an odd-cycle component of $G-S$. A \emph{chain} is a sequence defined by
$T_0=S$, $R_0=Z$, and for $j\ge0$: choose $p_{j+1}\in R_j$ (so $p_{j+1}\ne w_j$ for $j\ge1$), let $w_{j+1}$ be the
$T_j$-neighbour of $p_{j+1}$, let $R_{j+1}$ be the path given by Lemma~\ref{lem:exchange} applied
to $(T_j,\,Y_j,\,p_{j+1})$ with $Y_0=Z$ and $Y_j=R_j+w_j$, and set $T_{j+1}=T_j-w_{j+1}+p_{j+1}$.
Write $W_j=\{w_1,\dots,w_j\}\subseteq S$ and $P_j=\{p_1,\dots,p_j\}$, so $T_j=(S\setminus W_j)\cup P_j$.

\begin{lemma}[chain invariant]\label{lem:chain}
For every chain and every $j\ge1$:
(I1) $T_j$ is a tight set of maximum size and $Y_j=R_j+w_j$ is an odd-cycle component of $G-T_j$;
(I2) $R_j$ is a component of $G-S$, distinct from $Z,R_1,\dots,R_{j-1}$;
(I3) every vertex of a component of $G-S$ other than $Z,R_1,\dots,R_j$ has the same neighbours
in $T_j$ as in $S$; in particular the $S$-neighbours of a vertex of $R_j$ are its unique
$T_j$-neighbour together with $w_j$ if it is an endpoint of $R_j$.
\end{lemma}
\begin{proof}
Induction on $j$; (I1) is Lemma~\ref{lem:exchange}(c). Assume the claims up to $j$. The components
of $G-T_j$ are: (i) the components of $G-S$ other than $Z,R_1,\dots,R_j$; (ii) the path $Z-p_1$
and the paths $\rho_i=Y_i-p_{i+1}$, $1\le i<j$; (iii) the odd cycle $Y_j$.
Note $N(p_i)\subseteq R_{i-1}\cup\{w_{i-1},w_i\}$ and $N(w_i)=\{p_i\}\cup\{\text{endpoints of }R_i\}$.

\emph{$w_{j+1}\in S\setminus W_j$.} Otherwise $w_{j+1}=p_i$ with $i\le j$ would be adjacent to
$p_{j+1}\in R_j$; but $p_{j+1}\notin R_{i-1}$ by (I2) and $p_{j+1}\notin S$.

\emph{$R_{j+1}$ is of type (i).} It is a path, so not $Y_j$. It is not $Z-p_1$: its closing vertex
$w_{j+1}\in S$ would be adjacent to both neighbours $z^\pm$ of $p_1$ on $Z$, i.e.\ equal to both
of their (distinct, Lemma~\ref{lem:exchange}) $S$-neighbours. It is not $\rho_i$: if both endpoints
$y,y'$ of $\rho_i$ lie in $R_i$, then $w_{j+1}\in S\setminus W_j\subseteq T_i$ is the
$T_i$-neighbour of $y\in Y_i$ and is adjacent to $y'\in Y_i$, contradicting
Lemma~\ref{lem:exchange}(a) at stage $i$; otherwise one endpoint is $w_i\in S$, and $w_{j+1}\in S$
cannot be adjacent to it because $S$ is independent.

Hence $R_{j+1}$ is a component of $G-S$ distinct from $Z,R_1,\dots,R_j$, giving (I2). For (I3):
a vertex of a type-(i) component is not adjacent to any $p_i$ or $w_i$ by the neighbourhood
description above, so its $T_{j+1}$- and $S$-neighbourhoods coincide; the statement about $R_{j+1}$
follows from (I1) for $T_{j+1}$ (each vertex of $Y_{j+1}$ has exactly one $T_{j+1}$-neighbour) and
$T_{j+1}=T_j-w_{j+1}+p_{j+1}$.
\end{proof}

\paragraph*{Conclusion.}
Every chain can be extended: $R_j$ is non-empty, so some $p_{j+1}\in R_j$ exists, and
Lemma~\ref{lem:exchange} applies to $(T_j,Y_j,p_{j+1})$ because $T_j$ is a tight set of maximum
size (I1) and $G\ne K_4$. By (I2) the paths $R_1,R_2,\dots$ of an infinite chain are pairwise
distinct components of the finite graph $G-S$, a contradiction. Hence $\fr(G-S)=0$ for a maximum
tight set $S$, which is therefore an independent odd cycle transversal of size $\fr(G)$.

\paragraph*{Energy.} For $0<D_3\le1$ Theorem~\ref{thm:sandwich} and $\ioct=\fr$ give
$\OPT=D_3\fr$. For $D_3\ge1$, $\OPT\le\fr$ (no state 3) and
$E=m_{12}+e(S)+D_3|S|\ge m_{12}+|S|\ge\fr$ by the re-insertion argument. For $D_3=0$ the
independent odd cycle transversal gives a proper 3-colouring, $\OPT=0$. \hfill$\square$

\paragraph*{Remarks.} Planarity is not used. The argument does not fix a maximum cut, consistent
with the fact that not every maximum cut is healable (Appendix~\ref{app:numerics}). For planar $G$
the theorem makes $\OPT(D_3)$ computable in polynomial time for all $D_3$.

\subsection*{Constructive version}
Call a tight set $T$ \emph{augmentable} if $T+v$ is independent and tight for some $v$, and
\emph{stuck} otherwise. The tight sets of size $s$ are all stuck if and only if no tight set of
size $s+1$ exists (removing any vertex from a tight set of size $s+1$ leaves an augmentable tight
set of size $s$), so ``maximum size'' in the proof above is the statement that all tight sets of
size $|S|$ are stuck. The proof, however, uses this only for explicitly given sets: the current
set $T_j$ of a chain and the sets $T_j-t+y$ obtained by exchanging a vertex $y$ of the odd cycle
$Y_j$ with its $T_j$-neighbour $t$ (Lemma~\ref{lem:stuck} and Lemma~\ref{lem:exchange} applied to
them, including the exchanges at $y^\pm$ and at $z^\pm$ and the one used in the proof of (I2)).
This gives:

\begin{proposition}\label{prop:constructive}
Let $G\ne K_4$ be connected of maximum degree at most three, and let $S$ be a stuck tight set
with $\fr(G-S)\ge1$. Run the chain from an odd-cycle component $Z$ of $G-S$, and at each stage $j$
test all exchanges $T_j-t+y$, $y\in Y_j$. Within at most $c$ stages, $c$ the number of components
of $G-S$, one of these exchanges is an augmentable tight set of size $|S|$.
\end{proposition}
\begin{proof}
If none is augmentable, every set to which the proof applies maximality is stuck, so
Lemmas~\ref{lem:stuck}--\ref{lem:chain} and the conclusion hold verbatim with ``maximum size''
replaced by ``stuck'', and (I2) bounds the length of the chain by $c$, while the conclusion
extends it; a contradiction.
\end{proof}

Hence the following procedure constructs an independent odd cycle transversal of size $\fr(G)$:
starting from $S=\emptyset$, add a vertex whenever $S$ is augmentable, and when $S$ is stuck with
$\fr(G-S)\ge1$ move to an augmentable tight set of the same size by
Proposition~\ref{prop:constructive}. Each of the $\fr(G)$ rounds uses $O(|V|^2)$ evaluations of
$\fr$ on subgraphs of $G$, so on planar lattices the ground state itself, not only its energy, is
obtained in polynomial time. The exchange step is necessary: pure greedy augmentation gets stuck
with $\fr(G-S)\ge1$ on some graphs with 12 vertices (e.g.\ the planar graph with graph6 code
\verb+K?`D@`ae?iH_+ and $S=\{4,8,10\}$), although never on graphs with at most 11 vertices.

\begin{corollary}\label{cor:one-ground-state}
Given one optimal 2-colouring (Ising ground state) of $G$, an independent odd cycle transversal of
size $\fr(G)$, and hence a ground state for every $D_3$, is obtained in polynomial time without
any further maximum-cut computation.
\end{corollary}
\begin{proof}
Maintain a tight set $S$ together with an optimal colouring $c$ of $G-S$, $k=\fr(G-S)$.
\emph{Augmentation or certificate.} From a monochromatic edge $uv$ follow the walk in the proof of
Lemma~\ref{lem:stuck}, recolouring as there, which keeps $c$ optimal. Either it meets a vertex $x$
without neighbour in $S$; $x$ is an endpoint of a monochromatic edge of $c$, so $S+x$ is tight
(Lemma~\ref{lem:aug}), and $c$ restricted to $G-S-x$ has at most $k-1$ monochromatic edges while
$\fr(G-S-x)\ge k-1$ by (F1), so it is optimal. Or the walk closes and certifies that the component
is an odd cycle all of whose vertices have a neighbour in $S$. If every frustrated component is
certified, $S$ is stuck: every other vertex either has a neighbour in $S$ or lies in a bipartite
component, whose removal does not lower $\fr(G-S)$.
\emph{Exchange.} For $T'=T-t+y$ as in Lemma~\ref{lem:exchange}, colour the path $Y-y$ properly and
give $t$ the colour opposite to the majority of its at most two neighbours in $G-T'$. This colouring
of $G-T'$ has at most $(k-1)+1$ monochromatic edges and is optimal because $T'$ is tight.
Thus the procedure after Proposition~\ref{prop:constructive} runs on colourings alone; each walk,
exchange and certificate takes linear time, so the total time is polynomial.
\end{proof}

For planar lattices the Ising ground state is a minimum $T$-join; the healing step itself then
takes about one second for $4\times10^3$ sites in our implementation, a small fraction of the
matching time (Appendix~\ref{app:numerics}). Unlike the 2-SAT certificate of Sec.~\ref{sec:healing}, which may fail for a particular Ising
ground state, this procedure always succeeds.